\documentclass[revtex4]{emulateapj}
\usepackage{hyperref,graphicx}

\begin{document}
\received{1 april 2015}
\revised{22 may 2015}
\accepted{26 may 2015}

\newcommand{\fu}{FU~Ori}
\newcommand{\vten}{V1057~Cyg}
\newcommand{\vfif}{V1515~Cyg}
\newcommand{\vsix}{V1647~Ori}
\newcommand{\lkh}{LkH$\alpha$~188/G4}
\newcommand{\tts}{T Tauri star}
\newcommand{\kms}{km\,s$^{-1}$}
\newcommand{\Msun}{M_\odot}
\newcommand{\msun}{M$_\odot$}
\newcommand{\rsun}{R$_{\sun}~$}
\newcommand{\lsun}{L$_{\sun}~$}

\title{A Simple Calculation in Service of Constraining the Rate of FU~Orionis Outburst Events from Photometric Monitoring Surveys}
\author{Lynne A. Hillenbrand\altaffilmark{1}, Krzysztof P. Findeisen\altaffilmark{1,}\footnote{current address: Observatoire de Paris, CNRS UMR 8111 / G\'{E}PI, 5 Place Jules Janssen, 92190 Meudon, France}} 
\affil{$^1$Department of Astronomy; MC 249-17; California Institute of Technology; Pasadena, CA 91125, USA; \href{mailto:lah@astro.caltech.edu}{lah@astro.caltech.edu}}

\begin{abstract}
An enigmatic and rare type of young stellar object is the FU~Orionis class.  The members are interpreted as ``outbursting," that is, currently in a state of enhanced accretion by several orders of magnitude relative to the more modest disk-to-star accretion rates measured in typical T Tauri stars.  They are key to our understanding of the history of stellar mass assembly and pre-main sequence evolution, as well as critical to consider in the chemical and physical evolution of the circumstellar environment -- where planets form.  A common supposition is that {\it all} T Tauri stars undergo repeated such outbursts, more frequently in their earlier evolutionary stages when the disks are more massive, so as to build up the requisite amount of stellar mass on the required time scale.  However, the actual data supporting this traditional picture of episodically enhanced disk accretion are limited, and the observational properties of the known sample of FU~Ori objects quite diverse.  To improve our understanding of these rare objects, we outline the logic for meaningfully constraining the rate of \fu\ outbursts and present numbers to guide parameter choices in the analysis of time domain surveys.
\end{abstract}

\maketitle

\section{Introduction}

Pre-main-sequence stars are highly variable due to mechanisms 
operating in the stellar photosphere, in the magnetosphere and
innermost regions of the circumstellar disk, or in the disk atmosphere.
While the phenomenology 
has received attention since the work of Joy (1945) 
and was pioneered in the CCD era by Herbst et al. (1994), 
it is only in the last several years that large scale, quality, multi-year duration 
and moderate-cadence (e.g. Findeisen 2015; Parks et al. 2014, Rice et al. 2012)
or multi-week and high-cadence (e.g. Cody et al. 2010, 2011, 2014) 
sampling of the photometric time series phase space has become commonplace.  
These recently available data sets have enabled more complete characterization 
of the diversity of young star variability than was possible earlier,
and they serve to further our understanding of physical processes associated with young stars.

\begin{figure}[t]
\includegraphics[width=0.52\textwidth,angle=0]{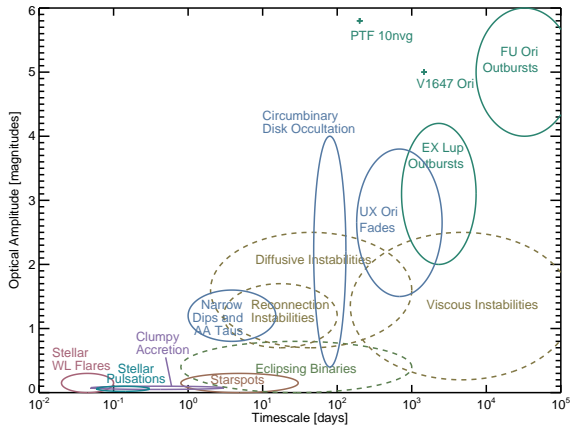}
\caption{ \small 
The variability parameter space occupied by empirically recognized phenomena 
(solid curves) and theoretically hypothesized mechanisms (dashed curves)
that operate in young stars.  
Some phenomena produce relatively symmetric flux variations (e.g. starspots), 
while others produce fading events (e.g. eclipses, variable extinction) or bursting events
(accretion variations).
This representation is intended to be schematic 
only, with any detailed use requiring quantification of the definitions of
``amplitude" and of ``time scale".  The FU~Ori stars occupy the upper right 
of the diagram. 
\label{fig:timeamp}}
\end{figure}

A number of identified or hypothesized variability mechanisms are illustrated 
in Figure~\ref{fig:timeamp}. Both periodic and aperiodic phenomena 
are represented, with short (minutes and hours) to long (years and decades) 
duration processes causing a wide range of brightness changes. 
Recent monitoring programs have contributed to the quantification 
of variability types, amplitudes, and timescales.  But depending on survey
parameters such as cadence, duration, and photometric precision, 
only some of the plausibly operating physical phenomena are identifiable 
in any given time series data set.  Thus, the relative distribution 
of young stars in the plane 
of amplitude vs time scale shown in Figure~\ref{fig:timeamp}, is unknown. 


\subsection{Typical Young Star Variability Patterns}\label{typical}

At lower amplitudes and intermediate time scales, significant progress
has been made in understanding observed young star variability phenomena.
Simple periodic modulation is attributed to rotation of 
photospheric inhomogeneities across the pre-main sequence stellar disk.  
For ``cool starspots" the analogy is to similar phenomena on the much older Sun,
but with significantly enhanced amplitudes (up to $\sim 0.1$ mag).
In addition, there may be rotationally modulated ``hot starspots"
attributed to accretion stream footprints.  
Variability of a periodic nature has been well studied 
over the past several decades, mostly in connection with a desire 
to measure angular momentum evolution and its relation to disks and accretion.

By contrast, aperiodic 
variability of a stochastic or perhaps only short time coherence nature, 
is less well-understood. This is in large part due to the lack of obvious
correlation between specific physics and specific variability patterns,
though aperiodic behavior is generally associated with circumstellar 
rather than stellar phenomena.  
Disk presence, accretion, and outflow are implicated.  Variability
amplitudes can exceed 1 mag but the typical rms is about 0.1 mag 
(e.g. Grankin et al. 2007, 2008 and Venuti et al. 2015 for optical,
Carpenter et al. 2001, 2002 for near-infrared; and 
Morales-Calderon 2011 for mid-infrared study).
Findeisen et al. (2013) identified moderate amplitude, 
intermediate time scale fading and brightening categories in optical data, 
with fades lasting weeks to years and likely caused by 
obscuring material in the few tenths to few AU range, and 
bursts lasting days to months and likely arising 
in disk processes that are also outside of the magnetospheric region.


Recently, Cody et al.\ (2014), Stauffer et al.\ (2014, 2015), and
McGinnis et al. (2015)  published high quality lightcurves 
of young stellar objects with disks in the young cluster NGC 2264.  
The combination of precision, cadence, continuous duration, 
and wavelength coverage, revealed exquisitely detailed 
photometric fluctuations that were used to identify
the following broad categories of variability: 

\begin{itemize}

\item
Flux dips with periodic or quasi-periodic nature are interpreted as 
circumstellar dust passing through our line of sight 
as it orbits, probably near the inner disk.  The dust is hypothesized
in different scenarios as either entrained in accretion or wind flows 
between the inner disk radius and the stellar photosphere (explaining ``narrow dip" stars)
or located in warps in the inner disk regions (explaining ``broad dip" 
stars a.k.a. AA~Tau analogs).
Flux dips that are aperiodic but repeating also likely related to dust in the
accretion flow.
\item
Brightening events relative to a more steady (but still variable)
flux level are intepreted as unsteady mass transfer from inner disks to stars.
The accretion flow has either variable flux along, or variable 
penetration of, a roughly dipolar magnetosphere.
These events are aperiodic and there is no corresponding periodic category.
\item
Phenomena that are rooted in the stellar photosphere include 
periodic modulation due to temperature inhomogeneities (cool/hot spots) 
that rotate with the photosphere.  Aperiodic events include
discrete flares due to coronal-like magnetic activity.
\end{itemize}

The recent plethora of monitoring data is 
enabling more detailed and quantitative study of the light curve shapes
and the variability time scales (e.g. Findeisen et al. 2015) and amplitudes.
The overall direction of work in the field now, is to tie {a mathematical description of}
typical variability patterns among young stars more directly to the driving physics.  

\subsection{Rare Burst and Rarer Outburst Behavior}\label{burst}

Young star bursts and outbursts are indicated in Figure ~\ref{fig:timeamp}
with amplitudes above $\sim$1 mag and time scales longer than $\sim$1 week. 
The statistics regarding these phenomena currently are quite poorly constrained.  

Young stars that undergo repeated outbursts with $\sim$2-4 mag amplitudes 
and characteristic burst timescales of weeks to months 
are called EX Lup type\footnote{These objects are sometimes called 
``EXor type variables" but they are actually named after EX Lup. 
The EXor notation that has crept into the literature is unfortunate 
since there is in fact a star EX Ori, but it is a pulsating AGB star.} 
variables (Herbig et al. 2001; Herbig 2008; Lorenzetti et al. 2009).
\vsix\ type outbursts are also repeating, with larger $\sim$4-5 mag amplitudes, 
and similar or perhaps longer characteristic timescales of up to a year or so (Aspin et al. 2009, 2006).
Finally, the \fu\ stars undergo $\sim$4-6 magnitude brightness increases 
on time scales of a few months to years (Herbig, 1989)
and then decay over at least decades to centuries 
(empirically) or millenia (theoretically).  
Because of the long times scales, repeated bursts
have not yet been observed among individual members of the FU Ori class.

To explain the moderate-amplitude bursts, there are a number of flavors 
of instabilties and pulsational behavior in circumstellar disks that have been
suggested to lead to variable and possibly cyclic inward mass flow.  
Figure~\ref{fig:timeamp} also shows the relevant ranges in
amplitude and time scale for various models.
Magnetic reconnection instabilities are modelled by Lovelace et al. (1995) 
and Romanova et al. (2002), Rayleigh-Taylor instabilities investigated 
by Romanova et al. (2004), diffusive instabilities proposed 
by Goodson \& Winglee (1999) and Romanova et al. (2005),
and viscous instabilities advocated by D'Angelo \& Spruit (2010, 2012). 
The largest amplitude and longest time scale 
events are thought to be driven by inner disk instabilities 
and associated rapid mass accretion, as described in detail below.

The \fu\ outbursts are the most extreme and sustained, while the other burst categories
are less extreme but more frequent.  
Even the stochastically variable burst behavior 
mentioned above as ``typical," is also probably driven by nonsteady or unstable accretion
from the inner disk to the star. 
Thus there is likely a broad continuum of burst behavior.
Understanding the frequency of occurence of different types of 
accretion burst events is important for improving our understanding
of stellar mass assembly. However,
the balance of influence among these different
burst categories on pre-main sequence and disk evolution is currently unknown. 
For the lower amplitude and more frequent events, rates are currently unstudied. 
For the higher amplitude and less frequent \fu\ stars, the event rate has been 
estimated in various ways in previous literature as discussed below,
though numbers range over several orders of magnitude.


We focus here on the infrequent, well-defined, large amplitude burst events exhibited by
\fu\ stars. The difference between modelling these events and modelling the
related EX Lup stars (shorter duty cycle and lower amplitude)
and V1647 Ori type objects (shorter duty cycle but similar amplitude, with evidence of 
the contributions of both accretion enhancement and extinction reduction in their brightening)
is that the \fu\ stars can be identified for decades after their bursts, whereas the
other types of objects must be identified during the weeks to months to year that they are bursting. 
We leave such an effort to a possible future contribution.

\section{The FU~Orionis Class of Objects}

\subsection{Phenomenology}

The \fu\ stars are a very rare group of young pre-main sequence objects. 
The prototype \fu\ and the two other defining members of the class, \vten\ and \vfif\ 
have their outbursts summarized from an observational perspective by Herbig (1977), 
while Larson (1980), Herbig (1989), Hartmann \& Kenyon (1985, 1996) 
discuss more general issues.  
Vittone \& Errico (2005) and Reipurth \& Aspin (2010) present recent review material.
\fu\ itself, the prototype of the class, was
known as an irregular variable star associated with the $\lambda$ Ori region; 
it underwent a $\sim$6 mag outburst
in 1936 and has decayed only 0.5-1 mag (where the range indicates the
uncertainty in the peak magnitude) since this time.
\vten\ was discovered as emission line object LkH$\alpha$ 190 by Herbig (1958)
and was presumed a normal \tts\ until it underwent a 5.5 mag outburst 
in 1969 thus becoming the second known \fu\ star; its decay is much more
rapid, and it is now only 2 mag above quiescence.  \vfif, also a known variable
before its outburst, experienced a much slower rise taking a dozen years 
to reach peak flux around 1980, 4 mag above quiescence,
and like \fu\ decayed only very slowly.  
In each of the above cases, with the source luminosity increasing by factors 
of hundreds to thousands,  the \fu\ outburst illuminated a reflection nebula. 

In addition to their outbursts,
the \fu\ objects are further distinguished among young stars, in that 
their optical spectra show unique signatures of early spectral type and 
low photospheric density, similar to F or G supergiants (Herbig 1977) 
while near-infrared spectra suggest M supergiants 
(Mould et al.\ 1978, Sato et al.\ 1992, Greene et al.\ 2008). 
The spectra are broadened by 
several tens to roughly one hundred \kms, much faster than the
typical absorption line widths of accreting T Tauri stars.  \fu\ stars 
also show spectral features and profiles associated with inflow and
outflow processes, such as classical P Cygni line profiles 
in H$\alpha$ and purely blueshifted absorption in higher Balmer lines, as well as
similar signatures in He, Na D, Ca II H\&K, and KI, attributed to winds.  
X-ray emission is also detected (Skinner et al. 2006).
The broad band SEDs of \fu\ stars (e.g. Gramajo et al. 2014)
exhibit large infrared excesses due to circumstellar material that is 
both thermally emitting and viscously dissipating energy from rapid accretion. 
Important for the disk instability interpretation is the finding by
Gramajo et al. (2014) that 80\% of \fu\ disks have $M_{disk}>0.1 M_\odot$,
more massive than any of the disks among the well-studied sample of T Tauri 
stars in Taurus, and that 90\% have accretion rates 
$dM_{disk}/dt > 10^{-6}~M_\odot$ year$^{-1}$, higher than 95\% 
of Taurus T Tauri stars, though for any individual source 
one of course must be cautious of reported model parameters.

The current census of \fu\ objects numbers only two dozen at most,
some still debatable.  



\subsection{Physical Interpretation}

To explain the high-amplitude outbursting \fu\   stars, theoretical models 
invoking thermal, 
gravitational\footnote{Though not widely recognized, the disk instability model
for \fu\   stars was first proposed by Paczynski in 1975, as 
reported by Paczynski, 1978 and Trimble, 1976.}, 
or magnetorotational disk instabilities, and combinations thereof, are proposed 
(Bonnell \& Bastien 1992; Bell \& Lin 1994; Clarke \& Syer 1996;
Kley \& Lin 1999; Armitage et al. 2001; Vorobyov \& Basu 2005; 
Boley et al. 2006; Zhu et al. 2009; Martin \& Lubow 2014). 
There are also suggestions of instability triggering 
through dense cluster or close binary or planet interactions 
(e.g. Pfalzner 2008; Reipurth \& Aspin 2004; Lodato \& Clarke 2004).  

The \fu\   broad-band SEDs are well-reproduced by the
disk models of Zhu et al. (2007, 2008) in which an outburst is driven by 
gravitational instability and associated magnetorotational instability that
extends out to several AU at an accretion rate of $10^{-5}~M_\odot$ year$^{-1}$
(Zhu et al. 2010).  
For the thermal instability mechanism (Bell \& Lin 1994), by contrast,
the outbursting zone is confined to $<$0.1 AU at $10^{-5}~M_\odot$ year$^{-1}$
but could reach 1 AU if $10^{-3} M_\odot$ year$^{-1}$; 
the SED fits of this model are less good  (Bell et al. 1995), 
although the light curves are well matched.  The enhanced 
mid-plane disk accretion rates can be compared to those typically inferred 
for classical T Tauri stars which are more like
$10^{-8}~M_\odot$ year$^{-1}$ (Gullbring et al. 1998).

In the standard picture (e.g. Hartmann \& Kenyon 1996), 
accretion outbursts are expected to decline in frequency and perhaps amplitude
with increasing source age.  More recent studies of protostellar accretion history 
(Vorobov \& Basu 2010; Stamatellos et al. 2012,
Dunham et al 2014; Bae et al. 2014)
include disk instability physics in various ways in simulations, which results
in the promotion of fragmentation. 
Accretion burst behavior can be produced 
at ages typically well less than $\sim$1 Myr, in the form of short-lived
1.5 to 2 orders of magnitude enhancement in disk accretion rates. 
Padoan et al. (2014), conversely, advocate 
for the importance of variable infall (rather than disk accretion) rates, which 
are prolonged in their turbulent cloud fragmentation model relative to
the more isolated core / envelope / disk models that rely on disk instabilities
to build up the stellar mass.

It is widely assumed that \fu\ stars are typical solar-type young stellar 
objects with central star masses $\sim$0.3-1.5 $M_\odot$ and ages $<$1-2 Myr. 
The predominant view is that they are Class I type sources, with a significant 
amount of the luminosity in their quiescent stages coming from envelope infall.
Some \fu\ stars, however, are demonstrated Class II type sources such as
PTF 10qpf (Miller et al. 2011), which was a routine M3e classical T Tauri star 
with little or no evidence based on the SED for envelope infall
\footnote{The Gramajo et al. (2014) study identifies 1/3 of known \fu\ stars 
as Class II, but detailed examination shows likely errors in both directions.
Most of their Class II designations (BBW~76 excepted)
have flat or rising SEDs, as expected for Class I and so-called flat-spectrum
sources, and some of their Class I designations (e.g. HBC 722 = PTF 10qpf) 
include far infrared photometry that is significantly confused 
by other sources in the beam.}.
Empirically, therefore, the episodic accretion behavior appears
to continue into the more optically revealed, envelope-free phase 
of star formation and pre-main sequence contraction, 
so possibly beyond 1 Myr of age.

\section{Estimating Outburst Rates}

The frequency of an astrophysical phenomenon is often derived 
by considering: the number of detections of a particular category of object, 
the biases for or against detection as well as the total number 
of sources in a parent sample from which the categories could be established, 
information about a relevant time scale.  With these values in hand one can
arrive at a rate of occurence.  A well-understood numerator 
and a similarly well-defined denominator are required for rigorous statistical 
assessment.  

\subsection{Previous Approaches to FU~Ori Outburst Rates}

Herbig (1977) advocated from the relative number of \fu\ stars and
regular T Tauri stars known at that time, that the \fu\ outburst rate was
$10^{-4} \textrm{ year$^{-1}$ star$^{-1}$}$, and that the phenomenon 
was recurrent. 

Hartmann \& Kenyon (1996) estimated the \fu\ outburst frequency
assuming the Miller \& Scalo (1979) star formation rate within 1 kpc 
of the Sun, i.e. 0.01-0.02 \msun\ per year, and also seem to assume an average 
mass close to solar, giving $\sim$0.01 new low mass stars per year.
For the census at the time of 5 known outbursts 
(FU~Ori, V1057~Cyg, V1515~Cyg, V1735~Cyg, and V346~Nor) in 60 years 
within 1 kpc, an integrated outburst rate of 0.08 per year is calculated.  
Dividing one rate by the other, a typical low mass star thus has $\sim$10 
outbursts.  Dividing the T Tauri disk lifetime of $1-10 \times 10^6$ years 
by 10 outbursts gives one outburst every $10^{5-6}$ years.  

A related McKee \& Offner (2010) estimation assumes: the Fuchs et al. (2009) 
star formation rate within 1 kpc of the Sun, 
$8 \times 10^{-3}$ \msun\ per year, and an average mass of 0.5 \msun, 
to get 0.016 low mass stars per year capable of outburst.  These authors 
also suggest that 25\% of the mass of a star is accreted through such
enhanced accretion episodes.

An updated census of outbursting objects by Greene et al. (2008) 
using the catalog of Abraham et al. (2004) tabulates 18 spectroscopically
identified \fu\ stars within 1 kpc; Reipurth \& Aspin (2010) provide a more recent compilation.
Dividing the number of presently known \fu\ objects by the star formation rate
above, 
a typical low mass star spends $\sim$1250 years of its life in an \fu\ phase.
Although consistent with the Hartmann \& Kenyon figure for a 1-century 
outburst length, the coincidence probably isn't meaningful since the increase 
in the \fu\ census in the last 15 years is offset by the higher 
star formation rate assumed.

The above statements are equivalent to saying that the rate of FU~Ori 
outbursts in active star forming regions is approximately $10^{-5} \textrm{ year$^{-1}$ star$^{-1}$}$, independent of whether all T~Tauri stars go through such a phase or whether only a subset is prone to outbursts. However, since the population of FU~Ori stars is incomplete, this rate is likely an underestimate.  

Another line of reasoning is that a typical T~Tauri star must have a lifetime average accretion rate of $\sim 10^{-6}~\Msun \textrm{ year}^{-1}$ to account for its estimated mass and age.  This high average accretion rate combined with the 
consequently lower-than-expected luminosity of ``protostars" led 
Kenyon et al. (1990) to argue that \fu\ stars were an important mechanism 
by which much of the time could be spent accumulating mass at low rates 
and relatively little time at higher rates.  If, for example, the star 
alternates between a quiescent phase of accreting $10^{-7}~\Msun \textrm{ year}^{-1}$ (higher than typical) and an outburst phase of accreting $10^{-4} \Msun \textrm{ year}^{-1}$ (also higher than what is thought typical), then one can account for all the star's accretion if it spends 1\% of its lifetime in outburst. Given an assumed typical outburst lifetime of $\sim100$~years, this implies an outburst rate closer to 
$10^{-4} \textrm{ year$^{-1}$ star$^{-1}$}$ is needed, or 
an even higher $10^{-3} \textrm{ year$^{-1}$ star$^{-1}$}$ if the more typical 
quiescent and outburst accretion rate values of $10^{-8}~\Msun \textrm{ year}^{-1}$ and $\sim 10^{-5}~\Msun \textrm{ year}^{-1}$, respectively,
are adopted.
Of note is that BBW 76 has been at maximum light since at least 1900 (Reipurth et al. 2002), and V883 Ori since at least 1888 (Cederblad 1946); similar to FU Ori, these stars are not rapidly fading.  Thus the actual duration of FU Ori outbursts may be a factor of several longer than what is typically assumed. 
Under the logic above, the outburst rate needed to reconcile the T~Tauri accretion rate vs age problem using the episodic FU~Ori outburst model is inversely proportional to the assumed outburst lifetime.  Longer duration bursts mean lower rates and, conversely, if 
 we allow for some fraction of shorter duration FU~Ori outbursts with lifetimes of only $\sim 30$~years, 
 then the needed outburst rate is higher by factors of several than the numbers above.

Bae et al. (2014) have recently produced a distribution of accretion rates
during the infall and early disk accretion phases up to 1 Myr, based on
2D hydrodynamic models that include heating and cooling.  Roughly
26\% of the time spent at the infall rate of 
$\sim 10^{-5.75}~\Msun \textrm{ year}^{-1}$, 
64\% of the time spent at the quiescent disk rate of 
$\sim 10^{-7.75}~\Msun \textrm{ year}^{-1}$, 
and 10\% of the time spent in outburst at
$\sim 10^{-5}~\Msun \textrm{ year}^{-1}$ 
(their Figure 16).  This results in an average accretion rate of 
$\sim 1.5\times 10^{-6}~\Msun \textrm{ year}^{-1}$ over the first $<$1 Myr, 
more than enough to accrete the stellar mass using the simple logic above. 

The burst rate in these new Bae et al. (2014) models 
is significantly higher than in previous gravitational instability scenarios 
where, inferring from their Figure 3,
about 24 outbursts in the first 0.3 Myr and only 2 in the next 0.7 Myr 
are suggested from the more standard models, whereas from their Figure 10
the number of oubursts is higher by at least a factor of two.
Converting to an estimated \fu\ rate results in values around
$8 \times 10^{-4} \textrm{ year$^{-1}$ star$^{-1}$}$ 
in the Class I stage, when the envelope dominates the disk extending to
about 0.25 Myr by which point the disk mass dominates, then dropping to about
$3 \times 10^{-6} \textrm{ year$^{-1}$ star$^{-1}$}$ into the Class II stage.

Considering the results of the numerous lines of reasoning above, 
it is not an exaggeration to say that there is currently 
a factor of at least 30 range among existing estimates of the \fu\ outburst rate.

\subsection{How to Measure the FU Ori Rate Empirically}

\begin{figure}[t]
\begin{center}
\vskip-4truein
\includegraphics[width=0.80\textwidth]{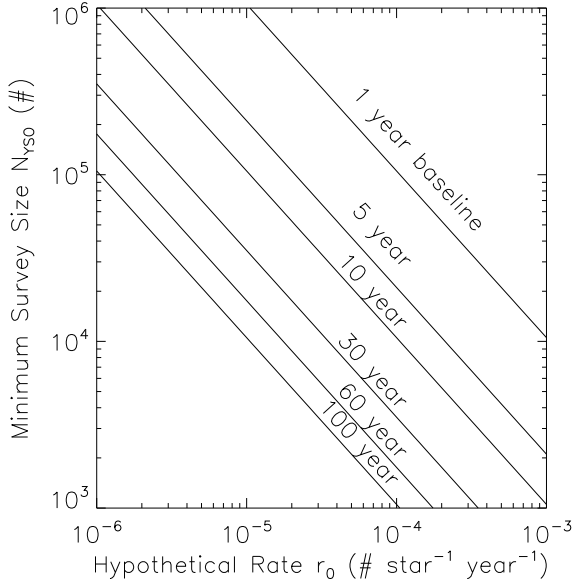}
\caption{ \small{
The survey size needed to provide a factor of two constraint on the FU Ori rate 
at 90\% confidence (C=0.9), with a 90\% chance of detecting enough events
to establish the constraint (R = 0.9).  The minimum sample size is plotted
as a function of the hypothetical ``true" outburst rate $r_0$ for surveys
having time baselines from 1 to 100 years.
One may reduce the needed survey size by choosing a longer time baseline, 
by admitting higher uncertainty than a factor of two, 
or by requiring a lower confidence than 90\%.
}
\label{fig:mdotmodel}
}
\end{center}
\end{figure}

One can consider it a reasonable goal to determine the frequency of FU~Ori 
outbursts to within a factor of two.  In the calculation that follows, 
we evaluate an experimental design intended to constrain the rate 
to a factor of two or better at confidence $C$, with a chance of success 
(i.e., a measure of the risk in carrying out the experiment) $R$. 
Both $C$ and $R$ have values in the range $[0,1]$. We   
assume a region of sky containing a number $N_{YSO}$ of young stars with disks 
(massive disks are required in the standard instability scenario 
for the outbursts)
that is monitored over duration $\Delta t$. We further assume: that outbursts 
occur at some average rate $r_0 = 10^{-\alpha} \textrm{ year$^{-1}$ star$^{-1}$}$,
that outbursts do not repeat on observable timescales 
(i.e. less than a century, essentially a constraint on $r_0 \Delta t \ll 1$~star$^{-1}$), 
and that any outburst occurring in the monitored interval is detected. 

The number of observed FU~Ori outbursts is then well modeled by 
a binomial distribution 
\begin{displaymath}
P(k|N, p) = {N \choose k} p^k (1-p)^{N-k}
\end{displaymath}
which, in our case with probability $p = r_0 \Delta t$, becomes
\begin{displaymath}
P(N_{event}|N_{YSO}, r_0 \Delta t) = {N_{YSO} \choose N_{event}} (r_0 \Delta t)^{N_{event}} (1- r_0 \Delta t)^{N_{YSO}-N_{event}}
\end{displaymath}
The number of FU~Ori events that are expected to occur is 
$N_{event} = N_{YSO}~r_0~\Delta t$ and the number detected 
would be $N_{obs} = \epsilon N_{event}$, where we assume
$\epsilon = 1$ in what follows. 
The probability of $n_{obs}$ or more detections 
$P(N_{obs} \geq n_{obs})$ is given by the appropriate number of binomial terms, $\sum_{k = n_{obs}}^{N_{YSO}} P(k|N_{YSO}, r_0 \Delta t)$. 

We assume that, in practice, the estimated outburst rate $r$, in outbursts per star per unit time, will be inferred from the observations $N_{obs}$, 
for example using $r = N_{obs}/\epsilon~N_{YSO}~\Delta t$.
We assume that the uncertainty associated with the measurement $r$ will be evaluated using the confidence interval formalism. 
We then ask how large a survey must be to have a high probability of detecting enough outbursts 
(given fluctuations around the mean \emph{true} outburst rate $r_0$) so as to produce a tight confidence interval around the measurement $r$.

Given an observed number of events $N_{obs}$ and an inferred outburst rate $r$, an observer can define a confidence interval 
$(r^\textrm{lower}, r^\textrm{upper})$, 
where the lower and upper limits on the outburst rate $r$ are
set by the desired confidence level $C$ and by $N_{obs}$. 
For a symmetric confidence interval these limits satisfy the following equalities:
\begin{displaymath}
P(n_\textrm{obs} \le N_\textrm{obs}) \equiv
   B(N_\textrm{obs}, N_{YSO}, r^\textrm{upper} \Delta t)  =  \frac{1-C}{2} \label{lowlim} \\
\end{displaymath}
\begin{displaymath}
P(n_\textrm{obs} \le N_\textrm{obs}) \equiv
   B(N_\textrm{obs}, N_{YSO}, r^\textrm{lower} \Delta t)  =  \frac{1+C}{2} \label{highlim}
\end{displaymath}
where in general $B(n_\textrm{obs}, n, p)$ is the cumulative distribution function for a binomial distribution with $n$ trials of probability $p$ that realizes $n_{obs}$ events. 
$B$ is monotonic with respect to all three arguments: holding the other two arguments fixed, $B$ increases as $n_\textrm{obs}$ increases, 
decreases as $n$ increases, and decreases as $p$ increases.

For the experiment to have provided a factor of two constraint, a symmetric confidence interval around 
the nominal measurement $r$ lies entirely inside the interval $[r/2, 2r]$. Thus we need to consider
the case where $r^\textrm{lower} \ge \frac{1}{2} r$ and $r^\textrm{upper} \le 2 r$. 
Substituting into the two equations above 
and expressing $r$ in terms of $N_{obs}$, we want to design the experiment so that it satisfies, with probability $R$, the inequalities
\begin{displaymath}
B(N_\textrm{obs}, N_{YSO}, 2 \frac{N_\textrm{obs}}{\epsilon N_{YSO}})  \le  \frac{1-C}{2} \\
\end{displaymath}
\begin{displaymath}
B(N_\textrm{obs}, N_{YSO}, \frac{1}{2} \frac{N_\textrm{obs}}{\epsilon N_{YSO}})  \ge  \frac{1+C}{2}
\end{displaymath}
The adjustable parameters in this experimental setup are the sample size
$N_{YSO}$ and the monitoring interval $\Delta t$.

For $\epsilon = 1$, $B(N_\textrm{obs}, N_{YSO}, 2 \frac{N_{obs}}{\epsilon N_{YSO}})$ is a decreasing function of $N_\textrm{obs}$, while $B(N_\textrm{obs}, N_{YSO}, \frac{1}{2} \frac{N_{obs}}{\epsilon N_{YSO}})$ is an increasing function. Therefore, for reasonable values\footnote{For an example of \emph{un}reasonable values, let $N_{YSO} = 2$. The first condition can be satisfied only if $N_\textrm{obs} = 1$, but $B(1, 2, \frac{1}{4}) = 0.9375$, so the second condition cannot be satisfied if $C > 0.875$.} of $N_{YSO}$ and $C$, both these criteria will be satisfied for sufficiently high $N_\textrm{obs}$. For a factor of two constraint at $C = 0.90$, we typically need $N_\textrm{obs} \sim 8$ for a broad range of $N_{YSO}$.

All that remains is to find the probability that $N_\textrm{obs}$ will be high enough, as a function of $N_{YSO}$. Since $N_\textrm{obs}$ is drawn from a binomial distribution with $N_{YSO}$ trials and probability $r_0 \Delta t$ (\emph{not} the observed rate $r$ that appears in the preceding paragraphs), this can be done by choosing trial values of $N_{YSO}$, finding the minimum number of $N_\textrm{obs}$ that would give a factor of two constraint on $r$, and then checking whether the probability of actually observing $N_\textrm{obs}$ events, given the true rate $r_0$, is at least our risk tolerance $R$.
In other words, we predict the results of a future statistical analysis, 
while integrating over all possible experimental outcomes to which 
said analysis would be applied.
The expressions above are solved simultaneously for the two parameters 
$N_\textrm{obs}$ and $N_\textrm{YSO}$. The smallest value of $N_\textrm{YSO}$ 
for which a solution exists is the minimum survey size.


The trade-offs in parameter space between event rate $r$, 
survey baseline $\Delta t$, and minimum survey size $N_{YSO}$ 
are illustrated in Figure \ref{fig:mdotmodel}.
The survey size required to probe a given outburst rate
is approximately inversely proportional to the available baseline 
(i.e., the expected event count is approximately constant). 
A sample of $\sim$1 million young stars with disks is needed
in order to probe outburst rates $\sim10^{-5}$ year$^{-1}$ star$^{-1}$
within a 1 year survey.  For longer time baselines, the survey samples can be
smaller and still probe the same outburst rate -- only $10^5$ stars are 
needed for a decade survey and $10^4$ for a century baseline in order to
probe the same rate as above.  We note right away that typical studies today 
report information on, at best, several thousand accreting young stars, and so
can not yet even begin to meaningfully constrain the FU Ori rate 
on empirical grounds.

\begin{figure}[t]
\begin{center}
\vskip-4truein
\includegraphics[width=0.80\textwidth]{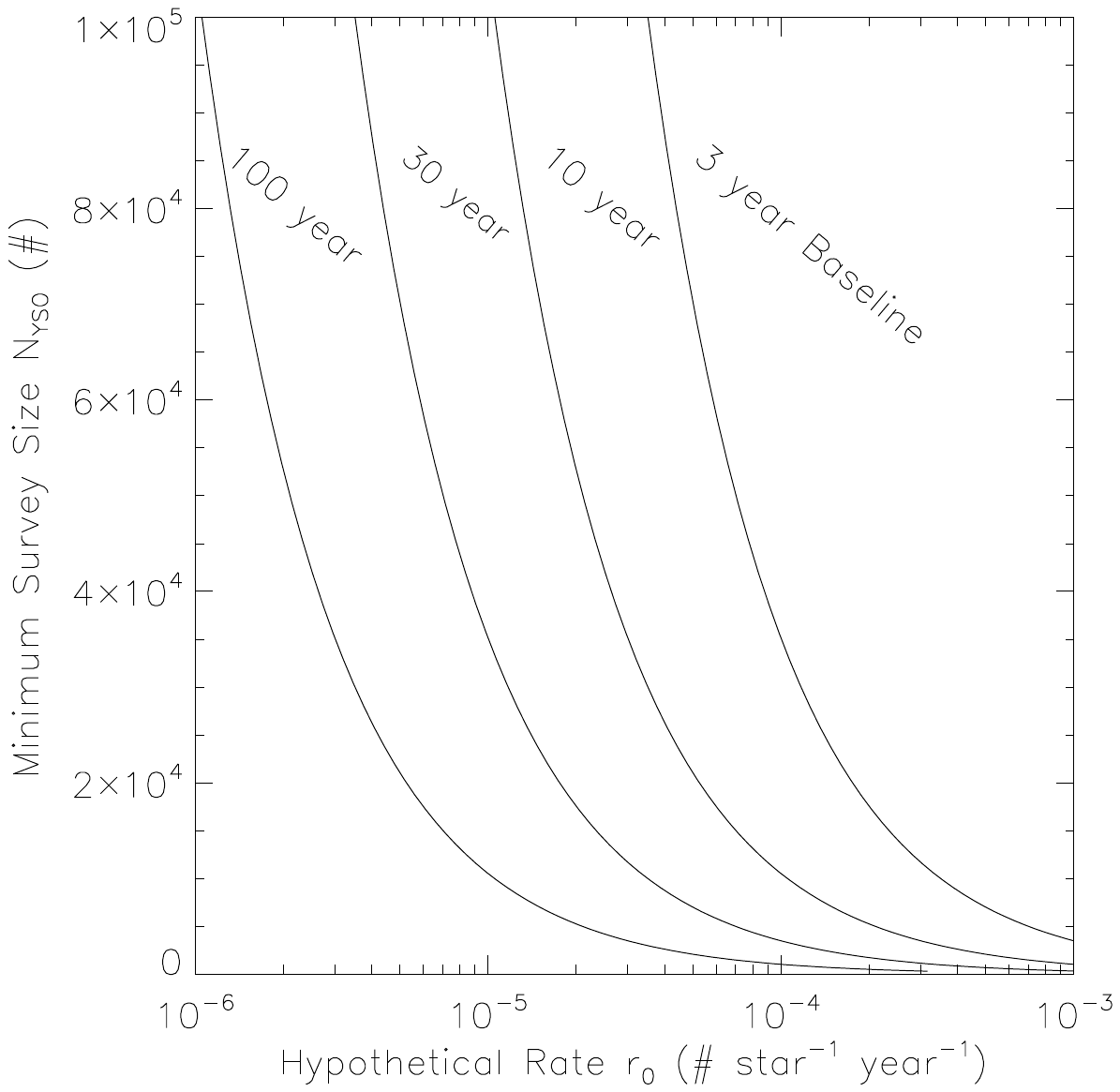}
\vskip-4truein
\includegraphics[width=0.80\textwidth]{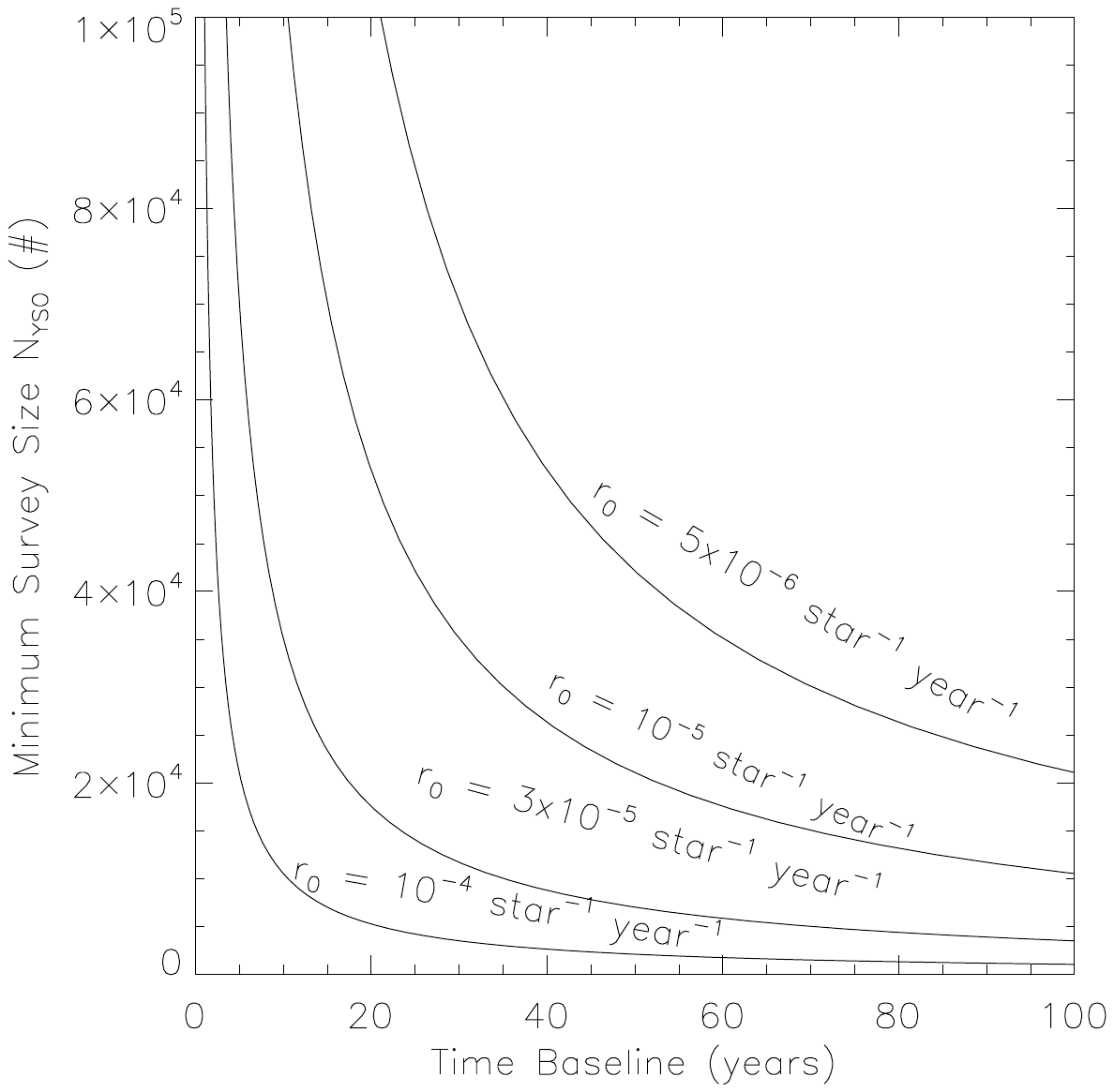}
\caption{ \small{
Cuts across Figure~\ref{fig:mdotmodel} that illustrate relationships among parameters
for survey sizes consisting of less than $10^5$ young stars with disks.
Top panel shows the outburst rate $r_0$ for various values of time baseline
$\Delta t$, while bottom panel shows the time baseline $\Delta t$ 
for various values of the outburst rate $r_0$ 
probed.  Both panels assume the same 90\% chance of constraining the outburst
rate to a factor of two. 
}
\label{fig:variations}
}
\end{center}
\end{figure}

Figure \ref{fig:variations} demonstrates the effect of considering
different time baselines and outburst rates for samples under $10^5$ stars,
a reasonable upper limit to the census of ``known" young stars within
a few kpc.
Samples larger than this are required if we aim to measure in a short survey
the true FU Ori outburst rate and it is much lower than
$10^{-4}$ year$^{-1}$ star$^{-1}$.

\begin{figure}[t]
\begin{center}
\vskip-4truein
\includegraphics[width=0.80\textwidth]{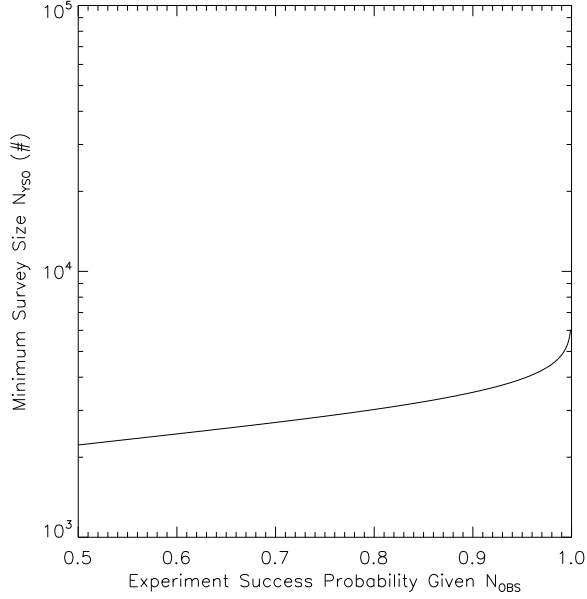}
\vskip-4truein
\includegraphics[width=0.80\textwidth]{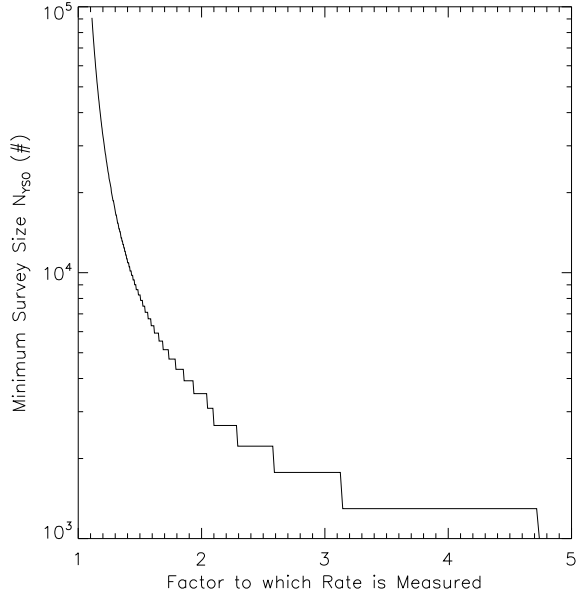}
\caption{ \small{
For a 30 year baseline and an outburst rate of 
$10^{-4}$ year$^{-1}$ star$^{-1}$ (i.e. $r_0 \Delta t = 3\times10^{-3}$), 
the survey size needed to detect a factor of two deviation from the hypothetical 
outburst rate, at 90\% confidence ($C=0.9$), and 90\% of the time ($R=0.9$).  
Required survey size is plotted as a function of (top panel) the probability $R$
that the survey will achieve the confidence bound, and (bottom panel) 
measurement precision on the hypothetical outburst rate.  
Larger surveys provide either higher probability of meeting specific precision
goals on the rate, or better precision on the rate.
Approximately 3500 stars are needed for factor of two precision 
on the rate (bottom panel) at 90\% confidence (top panel) 
for this time baseline and true outburst rate.
}
\label{fig:quality}
}
\end{center}
\end{figure}

Figure~\ref{fig:quality} illustrates the impact of altering some of the 
fixed parameters in the simulations above: the factor of two sensitivity 
to deviations from the particular outburst rate, 
and the required probability of a successful experiment. 
As expected, a larger survey, having smaller statistical fluctuations, 
provides either a higher probability of meeting specific precision goals
for the rate, or better precision on the rate.  The bottom panel shows that
relative to the factor of two constraint we have specified as adequate, improving our
knowledge of the FU Ori rate to a factor of 1.5 requires four times as much survey data
while knowledge to only a factor of 3 requires three times less survey data.
The top panel illustrates the relative insensitivity of the survey size requirement
to the risk of obtaining a less precise result on the rate than the experimental design. 
In other words, there is little to be gained by increasing the chance that a given
survey size will not detect enough FU Ori outbursts to obtain a reliable constraint on the FU Ori rate.

However, even if one sacrifices on quality of knowledge
(Figure~\ref{fig:quality}), it is impossible
to make useful statements about the \fu\ rate with a survey
of fewer than several thousand stars over fewer than several years
(Figures~\ref{fig:mdotmodel},~\ref{fig:quality}).
Considering larger samples,
for a survey of 10,000 stars, 
an outburst rate of $\sim10^{-3}$ year$^{-1}$ star$^{-1}$
can be measured or excluded after
a few years, which is a currently possible exercise,
while $10^{-4.5}$ year$^{-1}$ star$^{-1}$
can be probed after 100 years, also feasible now
by including particularly deep historical photographic plate data.
A total sample size above 10$^5$ stars would
systematically shift the lines to more rapid returns, as would relaxing
the requirements of 90\% confidence or factor of two knowledge of the rate.

\begin{figure}[t]
\begin{center}
\vskip-4truein
\includegraphics[width=0.80\textwidth]{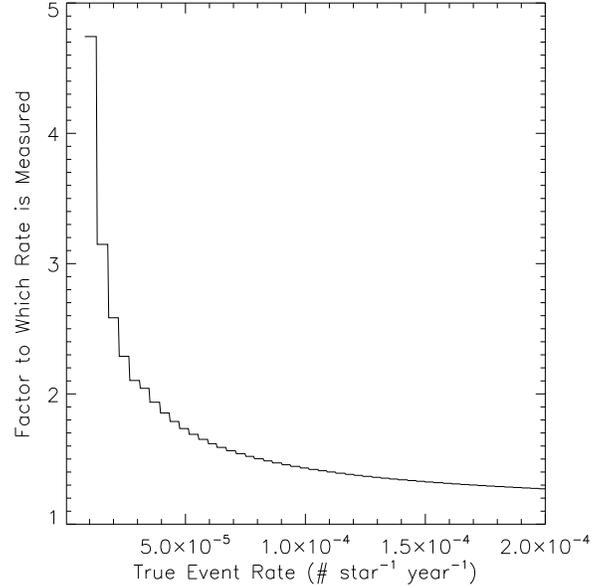}
\caption{ \small{
Dependence of our degree of knowledge of the FU rate on the rate itself.
Simulation results are shown for a survey 
of $10^4$ stars over a 30 year baseline 
(same as Figure~\ref{fig:quality}) assuming an experimental goal 
of obtaining a factor of two constraint 
on the FU Ori rate at $C=$90\% confidence, and that we tolerate an $R=$90\% chance 
of detecting enough sources to actually get the required constraint
(as elsewhere in this paper). 
At tip of the plot where the constraint is reduced from a factor of two  
to nearly a factor of 5, the probability of detecting at least one outburst 
for this sample size over this survey duration is exactly 90\%.
For higher event rates we would detect more outbursts and would improve our knowledge of the rate.
}
\label{fig:true}
}
\end{center}
\end{figure}

Finally, in Figure~\ref{fig:true} we explore the dependence of our ability to know
the FU Ori rate on what the event rate actually is.  As expected, the rarer the occurence
of FU Ori events, the looser the rate constraint from any given survey. 
The tip of the plot at low rates and poor degree of knowledge of the rate,
corresponds to a 90\% chance of detecting at least one FU Ori event
in the survey (in this case $10^4$ stars sampled over 30 years).

\section{Application to Modern Survey Data}

Increasing interest in young star variability mechanisms has led
to more and larger (both wider field and longer duration)
as well as often multiwavelength time series data sets. 
The value of the earliest possible first epoch observation is considerable, 
as illustrated in Figure \ref{fig:mdotmodel}.  Access to a 100 year baseline 
would mean either a reduction in the number of stars needed in the survey, or 
an improvement on the limit able to be placed on the \fu\ rate
-- both by the same factor of 100 -- relative to a 1 year baseline survey.

In practice, an automated comparison of two fields at widely separated epochs 
would select stars with measured magnitudes that differed by some 
threshold $\Delta m$ as candidate \fu\ objects, to be confirmed or rejected 
by follow-up observations. The actual magnitude increase $\Delta M$ 
to which such a strategy is complete is always larger than $\Delta m$ 
because of measurement errors in the photometry at each epoch.
Let a star have a true magnitude $M_1$ in epoch 1 and a true magnitude $M_2$ 
in epoch 2. Let the star have observed magnitudes $m_1$ and $m_2$ 
that can be modelled by Gaussian random variables with respective means
$M_1$ and $M_2$ and variances $\sigma_1^2$ and $\sigma_2^2$. 
The magnitude difference is then also a Gaussian random variable
with mean $M_2-M_1$ and variance $\sigma_1^2 + \sigma_2^2$. 
In this model, $m_2-m_1 > (M_2-M_1) - 1.282\sqrt{\sigma_1^2 + \sigma_2^2}$. 
with 90\% probability.  A survey with a cutoff $\Delta m$ for identifying 
an outburst detection will be 90\% complete to outbursts with amplitude 
larger than $\Delta m + 1.282\sqrt{\sigma_1^2 + \sigma_2^2}$. 

In an actual data analysis, the value of $\Delta m$ can be chosen arbitrarily 
based on the expected outburst amplitude and the quality 
of the available photometry. 
Statistical outliers based on photometric errors are a minor contaminant 
even if one targets only the weaker outbursts. For $\Delta M$ of 2 mag
and poor photographic photometry such as $\sigma_1 = 0.3$ mag, 
$\Delta m$ = 1.46 mag with a statistical false positive rate 
of $3 \times 10^{-4}$ star$^{-1}$. 
Raising the magnitude threshold, for example to exclude astrophysical 
false positives drawn
from the large variety of young star and other types variables that show 
shorter-term variations at the roughly 2 mag level, would effectively increase 
the sample size needed to ensure good statistics.

\subsection{Recent Results on High Amplitude Outbursts}

Attempting to address the FU Ori rate problem, 
Scholz (2012) used the near-infrared 2MASS-UKIDSS $\sim$8-year baseline 
to investigate several hundred stars 
and Scholz et al. (2013) used the mid-infrared Spitzer-WISE 5-year baseline 
to compare photometry of several thousand stars.  
Consequently, as clearly illustrated in Figure~\ref{fig:mdotmodel}, 
they could constrain the \fu\ rate to only much higher values
(by several orders of magnitude) than the historical estimates 
of the \fu\ rate that were discussed in \S{3.1}, 
established based on alternate considerations. 

From a survey with PTF consisting of over 1000 observations over the past 5.5 years 
towards the North America and Pelican Nebulae, two large-amplitude brightening
events were discovered:
PTF 10nvg (V2492~Cyg) which is likely a combination of accretion burst behavior 
and variable obscuration (Hillenbrand et al. 2013) perhaps analagous to V1647 Ori, 
and PTF 10qpf (V2493~Cyg) which is a bona fide \fu\ star 
(Semkov et al. 2010; Miller et al. 2011) previously known 
as the Class II classical T Tauri star \lkh or HBC 722. 
Both were discovered within the first season of PTF monitoring and -- given the
only few thousand star sample of known young stars in the region--  would
suggest an \fu\ rate significantly higher than the canonical value.

Considering a parent sample of $\sim$2000 stars in the region with evidence 
of circumstellar material, and assuming perfect detection efficiency, 
the implied event rate is then $9\times10^{-5}$ outbursts per star per year. 
From the simulations presented here, the 90\% confidence interval 
for this rate is [$3\times10^{-5},4\times10^{-4}$] outbursts per star per year.
Of course, statistical estimates based on only a single detection are not
adequate predictors.  
At the implied event rate, the chances of not seeing
another \fu\ outburst for another 20 years are only 3\%, while at the lower bound, 
there is a 30\% chance of no additional outbursts in the next 20 years. 

\subsection{Recent Results on Low Amplitude Bursts}

In addition to the bona fide \fu\   star outbursts, there are also 
lower-amplitude bursts that have been identified, including the EX Lup group
illustrated in Figure 1, as well as even shorter timescale and lower amplitude
bursts that are more routinely found in modern survey data.  
These less extreme but more common types may also play an important role 
in the accretion history of a star.

Among a sample of young stars selected based on variability 
and infrared-excess, Findeisen et al. (2013) found 
in the North America and Pelican Nebulae region that 12\% $\pm$ 3\% 
exhibited bursting behavior on timescales from 0.1 to 100 days, 
with amplitudes from 0.2 to 1.5 mag.  If confirmed as accretion-driven events,
the cases of repeated bursts reported there may be the first empirical examples 
of accretion-driven cyclic behavior caused by processes in the inner disk,
such as those predicted on time scales of a few tens of days 
in the disk instability mechanisms described in \S{2.2}.

The simulations presented above are not applicable 
to these shorter duration burst events since we have not included either the 
survey cadence (which needs to be short enough to catch and time-resolve
the burst) or the burst duration (which needs to be long enough to be detected)
as model parameters in our simple framework.



\subsection{The Numerator}

As stated earlier, estimating an event rate requires measuring
a number of detections, with considerations to bias, from among the
number of objects for which detections could be made.
The numerator in the rate equation is the number of \fu\ outbursts observed.
There are in fact two sides of the \fu\ outburst phenomenon which may be
amenable to detection: one the rise side or outburst itself,
occuring over weeks or months, and the other the decay side, occuring
over decades and centuries.

Numerous projects dedicated to other (usually extraglactic) science areas
are also finding young star outbursts, some of which are even \fu\ like.
The currently operating ASAS, CSS, and PTF will be succeeded by
next generation surveys such as ATLAS, ZTF, and LSST, all of which
are capable of making discoveries in this area.  The ATLAS project, 
for example, is predicting 5 \fu\ stars in its first year of operations.

Survey yield will depend on parameters such as sensitivity, cadence, 
and galactic latitude coverage, as well as on astrophysical factors 
such as the range in burst amplitudes and rise times, 
and the frequency of visibility at optical vs infrared-only vs submm-only 
wavelengths (e.g. Johnstone et al 2013).  
None of these considerations are included in our simple simulations, 
in which they were wrapped up in the factor $\epsilon$ which was set to unity.  
However, the results may be generalized to the case of $\epsilon \ne 1$
by considering that, even if not all \fu\ sources are detected, a 
Poisson/binomial process will still be observed albeit at an effectively 
reduced rate.  A best estimate rate and a confidence interval could still
be calculated, then multiplied by $1/\epsilon$. 

In addition to catching the \fu\ rise phase, various of these imminent 
time domain sky survey projects could also identify \fu\ objects 
during the decay phase, by selecting for follow-up spectroscopy 
sources with consistent fading over their 5-10 years of operations.  
However, there is significant diversity in the post-outburst light curves 
of known \fu\ stars (see e.g. Clarke et al. 2005, their Figure 1).  For example,
\fu\ itself has decayed only slightly in 50 years, 
while \vten\ had a much more rapid decay; \vfif\ was slower to rise, 
but has had a relatively flat post-outburst light curve.  
%

%


\subsection{The Denominator: Census of Selected Star Forming Regions}

The demoninator in the rate equation is the number of young stars
with the possibility to undergo an \fu\ outburst in a given survey.
The true survey size is $N_{YSO}$ young stars with disks, which given a certain
survey area is generally far less than as the number of objects 
photometrically detected in the field. Young stars are identified
via a variety of mechanisms associated with either stellar or circumstellar processes. 
Taken in isolation, none may not be 100\% reliable indicators of membership in $<$1-3 Myr old star forming regions,
but in combination they can lead to secure knowledge of the total size 
and identity of the targetted young stellar population.  

The usually considered diagnostics of young T Tauri stars are: 
photometric variability, low excitation emission lines,
presence of quick-burning lithium,
infrared excess due to circumtellar dust, 
ultraviolet excess due to re-radiated accretion shocks, 
and x-rays from both accretion and stellar coronal activity.   
As the \fu\   phenomenon is driven by massive disks, only so-called Class-I
and Class-II stars should be included in assessing survey statistics.
This is a much easier scenario than other statistical problems such as
the lifetime of disks where both the disked and non-disked populations
must be thoroughly known, as well as the absolute source ages.

There has been great progress over the last decade in terms of census building in star-forming
regions.   However, the job is not complete -- even for regions within only a few hundred parsecs of the Sun.
Spitzer-identified (infrared excess) and Chandra-identified (x-ray) samples are incomplete, even in combination, in part 
because these were pointed missions.  
The recent availability of wide-field survey data from WISE (mid-infrared),
2MASS and UKIDSS (near-infrared), iPHAS/UVEX (optical),
and GALEX (ultraviolet) can help fill in these census gaps near
and in the galactic plane. 
Non-outburst photometric variability is also -- after some decades
of hiatus -- again being used to identify young stars.  
GAIA results on kinematics will help define cluster and moving group membership 
for optically visible young stars.

\section{Discussion and Conclusions}

There are at present reasonable empirical constraints on the rate of occurence 
of phenomena such as M dwarf flares, novae, supernovae, and gamma ray bursts.
Each can be quantified in terms of the number per year per square degree 
per magnitude interval.  The young star \fu\  outbursts, however, which have
luminosity increases between those of M dwarf flares and novae,
are rare enough that only crude estimates exist 
for the number that occur within each young star disk lifetime. 
And while the rate is a significant number, more meaningful would be knowledge
of the distribution of time separation between \fu\ events during 
the first one to few Myr when the star first becomes visible as a 
self-luminous hydrostatically contracting object.

In this paper we have provided a simple framework for considering how well
we might be able to constrain the rate of \fu\ outburst events.  
The method employs binomial statistics which are generally applicable 
to any phenomenon that can be treated as either occurring 
(e.g. an outburst is detected) or not occuring (no outburst is detected).  
In our presentation we have provided numbers based on a 90\% measurement confidence 
requirement (a parameter of the statistical analysis that depends on the data 
actually obtained) with a 90\% probability (a parameter which refers to 
having a representative sampling of the astrophysics), under the assumption
that all outburst which occur are detected.
Figures 2, 3, 4, and 5 illustrate the simulation results.

With the rigor specified above, empirically constraining the \fu\ outburst rate
to a level of $<10^{-4}~\Msun \textrm{ year}^{-1}$ 
requires $>10^5$ young stars with disks to be monitored for $>1$ year,
or $>10^4$ such stars for $>10$ years.
Similarly, constraining the \fu\ outburst rate to 
$<10^{-5}~\Msun \textrm{ year}^{-1}$ 
requires $>10^5$ young stars with disks to be monitored for $>10$ years,
or $>10^4$ such stars for $>100$ years.  Finally
outburst rates $<10^{-6}~\Msun \textrm{ year}^{-1}$ could be probed 
if $>10^6$ young stars with disks were monitored for $>10$ years,
or $>10^5$ such stars for $>100$ years.
In any case, more than several $10^4$ suitably disked young stars 
are needed in order to be on the linear parts of these scaling relations,
given the likely range of true outburst rates.  
An extension of our calculation to shorter duration event types such as
EX Lup / V1647 Ori or lower level bursts would require consideration 
of the survey cadence relative to the burst duration as additional model parameters.

Finally, we note that in order to estimate the
true outburst rate, attention is needed not only to the numerator
that reflects the number of actually detected outburst events, but to careful definition
of the denominator, that is, the number of comparably aged young disked stars 
in the studied region.

\end{document}